\documentclass[a4paper,10pt]{article}

\usepackage[margin=25mm]{geometry}
\usepackage{times}
\usepackage{graphicx}
\usepackage{caption}
\usepackage{hyperref}
\usepackage{changepage}
\usepackage{todonotes}
\usepackage[nolist,nohyperlinks]{acronym}
\usepackage{amsmath}   
\usepackage{amssymb}   
\usepackage{amsfonts}  

\usepackage{fancyhdr}
\usepackage{titlesec}
\titleformat{\section}{\bfseries\large\uppercase}{\thesection.}{1em}{}
\titleformat{\subsection}{\bfseries\normalsize}{\thesubsection}{1em}{}
\titleformat{\subsubsection}{\itshape\normalsize}{\thesubsubsection}{1em}{}

\renewenvironment{abstract}{
  \begin{center}
    {\large\bfseries Abstract \par} 
  \end{center}
}{\par\vspace{2em}}

\usepackage[numbers]{natbib}

\newcommand{\prob}{\ensuremath{{p}}}

\newcommand{\data}{\ensuremath{{d}}}

\renewcommand{\exp}{\ensuremath{\text{exp}}}
\newcommand{\given}{\ensuremath{{\,\vert\,}}}

\begin{document}

\title{\bfseries Learned harmonic mean estimation of the marginal likelihood for multimodal posteriors with flow matching}
\author{Alicja Polanska$^{1}$, Jason D. McEwen$^{1,2}$ \\
{\small $^{1}$Mullard Space Science Laboratory, University College London, Dorking, UK}\\
{\small $^{2}$Alan Turing Institute, London, UK}}
\date{}
\maketitle
\thispagestyle{fancy}

\begin{abstract}
	The marginal likelihood, or Bayesian evidence, is a crucial quantity for Bayesian model comparison but its computation can be challenging for complex models, even in parameters space of moderate dimension. The learned harmonic mean estimator has been shown to provide accurate and robust estimates of the marginal likelihood simply using posterior samples. It is agnostic to the sampling strategy, meaning that the samples can be obtained using any method. This enables marginal likelihood calculation and model comparison with whatever sampling is most suitable for the task. However, the internal density estimators considered previously for the learned harmonic mean can struggle with highly multimodal posteriors. In this work we introduce flow matching-based continuous normalizing flows as a powerful architecture for the internal density estimation of the learned harmonic mean. We demonstrate the ability to handle challenging multimodal posteriors, including an example in $20$ parameter dimensions, showcasing the method's ability to handle complex posteriors without the need for fine-tuning or heuristic modifications to the base distribution.

\end{abstract}

\section{Introduction}
Bayesian parameter estimation and model comparison are essential tools in many areas of modern research. There exists a variety of methods that are widely used for these purposes, most notably \ac{MCMC} sampling methods \citep{metropolis:1953,hastings:1970} and nested sampling \citep{skilling2006nested}. As many fields, such as astrophysics, transition into a data-rich era driven by next-generation experiments, the scientific community has the opportunity to use these large datasets to evaluate highly complex and comprehensive models. However, the computational cost of performing Bayesian inference on these potentially high-dimensional models can become very large, motivating the development of accelerated and scalable inference methods. Traditional \ac{MCMC} methods can struggle with multimodal distributions and high dimensionality. Many scalable sampling methods have been developed recently, that leverage gradient information and modern accelerators (GPUs and TPUs) to explore the parameter space with high efficiency, such as \ac{HMC} \citep{neal2011mcmc,betancourt2015hamiltonian}, including \ac{NUTS} \citep{hoffman2011nouturn}, as well as machine learning-based or -assisted frameworks, such as \texttt{flowMC} \citep{wong2022flowmc, gabrie2022adaptive} or \ac{SBI} \citep{beaumont2019approximate,cranmer2020frontier}.

These methods excel at generating posterior samples in high dimensions but do not provide the marginal likelihood, a crucial quantity for Bayesian model comparison, like nested sampling does. On the other hand, nested sampling restricts the sampling strategy to be performed in a nested manner, as the name suggests. In previous work we introduced the learned harmonic mean estimator, which allows for accurate and robust estimation of the marginal likelihood simply using posterior samples \citep{mcewen2023machine,polanska2023learned,Polanska_2025}. While the original harmonic mean estimator \citep{newton1994approximate} is notoriously unstable due to a infinite variance \citep{neal:1994}, the new approach allows us to overcome this problem by leveraging machine learning. This allows for robust evidence calculation using samples generated by accelerated and scalable \ac{MCMC} methods \citep{piras2024future,polanska2024accelerated} and \ac{SBI} \citep{Spurio_Mancini_2023}. However, the previous implementation of the flow architecture used to learn the internal target density had limitations when dealing with highly multimodal posteriors. In this work we overcome these limitations by introducing a new architecture based on flow matching \citep{lipman2024flowmatchingguidecode}, which allows us to better capture complex posterior structures. Flow matching offers a simulation-free objective for training continuous normalizing flows, resulting in models that are significantly more expressive and easier to train than previously introduced architectures. We demonstrate the performance of this approach on a series of challenging examples with highly multimodal posteriors, including an example in $20$ parameter dimensions.

\section{Methodology}
\subsection{Learned harmonic mean estimator}
In the Bayesian framework, the central quantity of interest is the posterior distribution $\prob(\theta \given \data, M)$, the probability density of a model's parameter $\theta$ given observed data $\data$ and model $M$. It can be expressed using Bayes' theorem as:
\begin{equation}
	\label{eq:bayes}
	\prob(\theta \given \data, M)
	= \frac{\prob(\data \given \theta, M) \prob(\theta \given M)}{\prob(\data \given M)}
	= \frac{\mathcal{L}(\theta) \pi(\theta)}{z},
\end{equation}
where $\prob(\data \given \theta, M) = \mathcal{L}(\theta)$ is the likelihood, $\prob(\theta \given M) = \pi(\theta)$ the prior and $\prob(\data \given M) = z$ is the evidence.

The Bayesian evidence is a quantity of paramount importance in Bayesian model comparison. It measures the probability of observing the data under a particular model
\begin{equation}
	\label{eqn:evidence}
	z =
	\prob(\data \given M)
	= \int \,\text{d} \theta \
	\prob(\data \given \theta, M) \prob(\theta \given M)
	= \int \,\text{d} \theta \
	\mathcal{L}(\theta) \pi(\theta),
\end{equation}
measuring support for the model given the observed data. This allows for quantitative model comparison via the Bayes factor, which is the ratio of evidences for two competing models. However, computing the evidence is often challenging, especially in high-dimensional parameter spaces or for complex models, as it involves evaluating a multidimensional integral over the entire parameter space.

In recent work, the learned harmonic mean estimator of marginal likelihood was introduced \citep{mcewen2023machine}. It requires only samples from the posterior and their corresponding probability densities. In the Bayesian inference framework, this is often available from performing parameter estimation, and can be obtained using any method, including \ac{MCMC} and \ac{HMC} methods, such as \ac{NUTS}, as well as \ac{SBI} \citep{Spurio_Mancini_2023} and variational inference \citep{Blei_2017}. It builds upon the original harmonic mean estimator \citep{newton1994approximate} but overcomes its infinite variance issue by using machine learning methods to find a suitable normalised internal target distribution $\varphi(\theta)$. The reciprocal marginal likelihood $\rho = z^{-1}$ is then estimated as:
\begin{equation}
	\label{eqn:harmonic_mean_retargeted}
	\hat{\rho} =
	\frac{1}{N} \sum_{i=1}^N
	\frac{\varphi(\theta_i)}{\mathcal{L}(\theta_i) \pi(\theta_i)} ,
	\quad
	\theta_i \sim \prob(\theta | \data).
\end{equation}
The variance of this estimator is finite when the target density is contained within the posterior in parameter state. The target should be close in shape to the posterior itself, but crucially with thinner tails. This is challenging to achieve, especially in higher dimensions or for complex posteriors, particularly with many modes. In recent work normalizing flows were introduced to learn the target distribution \citep{polanska2023learned,Polanska_2025}. In contrast to simple machine learning methods used in previous work \citep{mcewen2023machine}, they do not require a bespoke training approach or hyperparameter tuning, making the method more robust and scalable. The flow is trained on posterior samples using forward \ac{KL} divergence, giving a normalised approximation of the posterior. To ensure the target has thinner tails than the posterior, the flow distribution is then concentrated by introducing a temperature $T \in (0,1)$, which is the scaling factor for the flow's base distribution variance. The improved robustness and scalability was demonstrated on practical applications within astrophysics, including a cosmological example in up to $159$ parameter dimensions \citep{piras2024future}, and a gravitational waves astrophysics example \citep{polanska2024accelerated}.

\subsection{Flow matching}
Discrete normalizing flows are limited in expressivity, and can struggle when dealing with complex, multimodal posteriors \citep{pmlr-v119-cornish20a}. When trained on samples from a multimodal distribution by forward \ac{KL} divergence, the flow tends to assign non-negligible probability mass to density bridges connecting distinct modes. For the learned harmonic mean estimator to have finite variance, it is critical that the target density $\varphi(\theta)$ is concentrated within the high-probability regions of the posterior and decays rapidly in the tails. The learned harmonic mean has been successfully applied to a highly multimodal gravitational wave example \citep{polanska2024accelerated}, but required heuristic modifications to the base distribution to accommodate the multimodality of the posterior. Specifically, a hand-picked Gaussian mixture was considered, to overcome the topology-preserving limitation of normalizing flows. The need for heuristics limits the flexibility and ease of use of the method. In this work, to address these limitations, we employ a flow matching-based continuous normalizing flow framework \citep{lipman2023flowmatchinggenerativemodeling,liu2022flowstraightfastlearning,albergo2023buildingnormalizingflowsstochastic}.

Flow matching is a training framework for \acp{CNF} that is more expressive and computationally scalable than traditional discrete flows. Unlike discrete flows, whose architectures are often constrained to keep Jacobians tractable, flow matching allows for the use of architecturally unrestricted, high-capacity neural networks.

\acp{CNF} parametrise the transformation between a simple base distribution, usually a standard Gaussian, and the distribution that is being approximated via a time-dependent velocity field \citep{NEURIPS2018_69386f6b,grathwohl2018scalable}. The approximate distribution can then be sampled from, and its density can be evaluated by solving the corresponding \ac{ODE}. In initial work on \acp{CNF}, training required expensive \ac{ODE} simulation and its differentiation during training. Flow matching uses a simulation-free regression objective to train the velocity field, making training highly efficient. Flow matching is scalable, and crucially for our applications, well-suited for multimodality \citep{albergo2023buildingnormalizingflowsstochastic}.

We consider samples from a posterior distribution of interest, that we denote as $p(\theta) \equiv \prob(\theta \given \data, M)$ and its approximating flow $p_{\text{NF}}(\theta, \beta)$, where $\beta$ are the trainable flow parameters. We define a probability path $p_t(\theta)$ that interpolates between a simple base distribution $\pi_{\text{base}}(\theta)$ (typically a standard Gaussian) at time $t=0$, and the complex target posterior $p(\theta)$ at time $t=1$. The flow is defined by an \ac{ODE} governed by a time-dependent velocity field $v_t(\theta)$:
\begin{equation}
	\frac{d\theta}{dt} = v_t(\theta, \beta).
\end{equation}
To train the network, we utilise the conditional flow matching objective with an optimal transport interpolant \citep{lipman2023flowmatchinggenerativemodeling}. We define a conditional probability path between a specific noise sample $\theta^0 \sim \pi_{\text{base}}(\theta)$ and a specific posterior sample $\theta^1 \sim p(\theta)$ as a straight line:
\begin{equation}
	\theta^t = (1 - t)\theta^0 + t\theta^1 .
\end{equation}
This path is generated by a simple, conditional vector field $u_t(\theta | \theta^1)$, which is the derivative of the path with respect to time:
\begin{equation}
	u_t(\theta | \theta^1) = \theta^1 - \theta^0.
\end{equation}

The neural network $v_t(\theta, \beta)$ is then trained to approximate the vector field by minimising the expected mean squared error over random time steps $t \sim \mathcal{U}(0, 1)$:
\begin{equation}
	\label{eq:loss}
	L(\beta) = \mathbb{E}_{t, \theta^0, \theta^1} \left[ || v_t(\theta^t, \beta) - (\theta^1 - \theta^0) ||^2 \right].
\end{equation}
It can be shown that, under mild assumptions, minimising this loss is equivalent to minimising the \ac{KL} divergence between the flow distribution $p_{\text{NF}}(\theta, \beta)$ and the target posterior $p(\theta)$ \citep{lipman2023flowmatchinggenerativemodeling}.
Once trained, the density $p_\text{NF}(\theta)$ required for the estimator is evaluated by solving the corresponding \ac{ODE} backwards from $t=1$ to $t=0$ while integrating the instantaneous change of variables formula:
\begin{equation}
	\log p_\text{NF}(\theta^1) = \log \pi_{\text{base}}(\theta^0) - \int_0^1 \text{div}(v_t)(\theta^t) dt.
\end{equation}
The divergence term can be found by accessing gradients of the velocity field with respect to $\theta$ using automatic differentiation, or approximated using Hutchinson's trace estimator \citep{hutchinson1989stochastic} for improved scalability.
\begin{figure}[t]
	\centering
	\includegraphics[width=10cm]{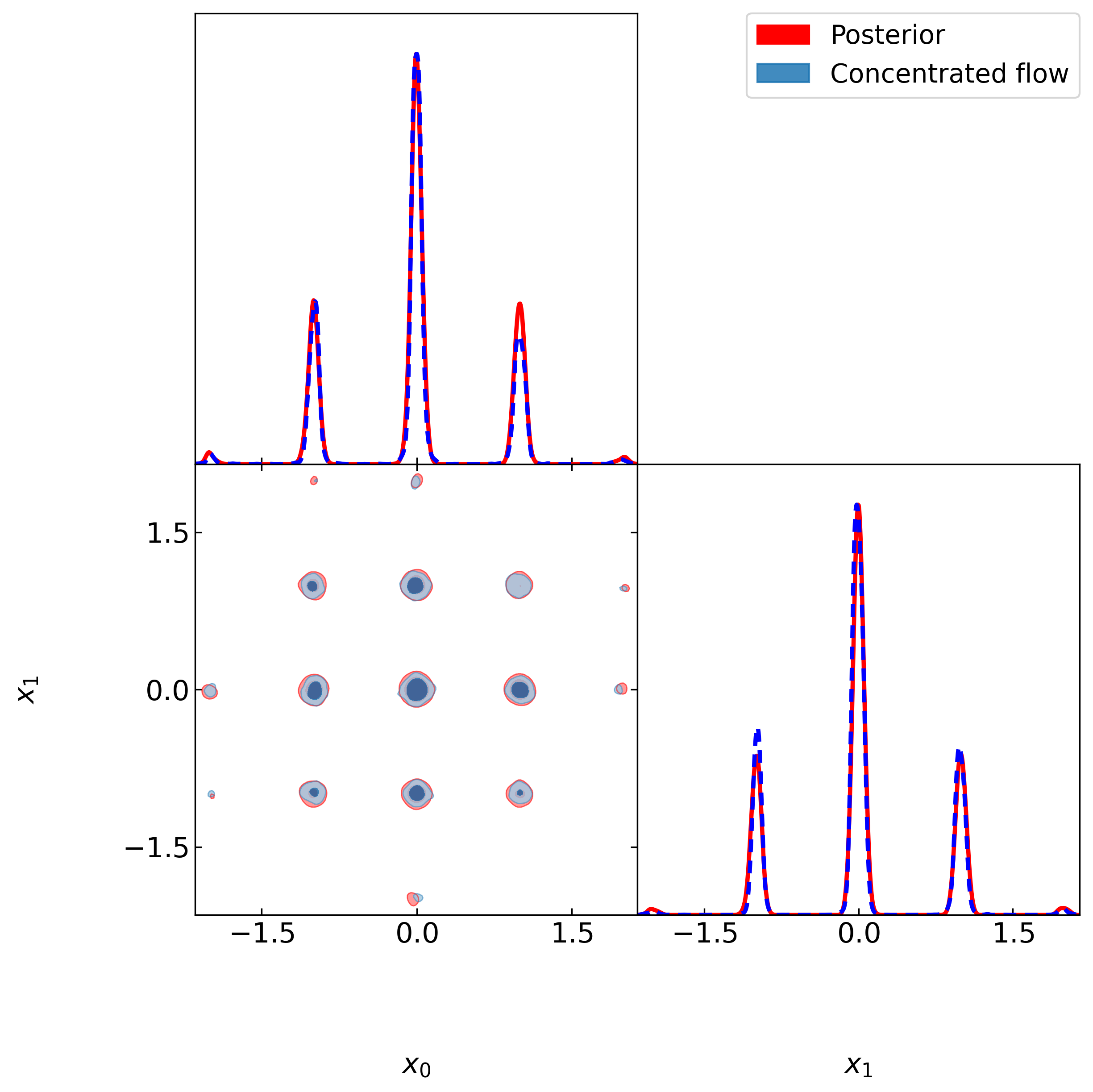}
	\caption{Corner plot of samples from the posterior (red) and trained flow at temperature $T=0.98$ (blue) for the Rastrigin example. The target distribution given by the concentrated flow is contained within the posterior as required for the learned harmonic mean estimator. All represented modes are captured by the flow accurately.}
	\label{fig:rastrigin_corner}
\end{figure}

\begin{figure}[ht]
	\centering
	\includegraphics[width=10cm]{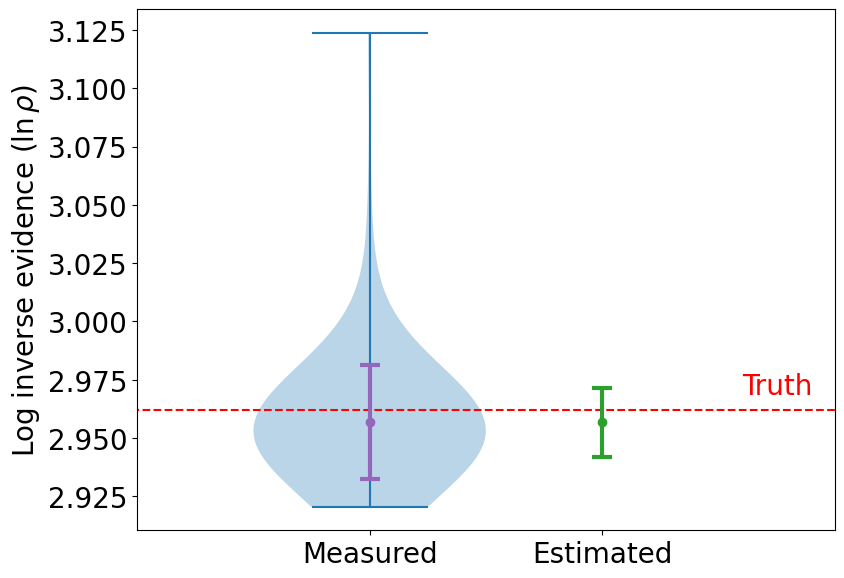}
	\caption{Marginal likelihood computed by the learned harmonic mean estimator for the Rastrigin example.
		Violin plot of $100$ runs of the experiment, showing the distribution of marginal likelihood values (measured) along with the error estimate computed by \texttt{harmonic} (estimated). The ground truth obtained directly by numerical integration is indicated by the red dashed line.}
	\label{fig:rastrigin_violin}
\end{figure}

\subsection{Learned harmonic mean with flow matching}
In this work we integrate the flow matching framework into the learned harmonic mean estimator to find the target distribution $\varphi(\theta)$. For the velocity field $v_t(\theta, \beta)$, we use a simple \ac{MLP} architecture, with the Swish activation function \citep{ramachandran2017swish}. The network is trained using the Adam optimizer \citep{kingma2017adam} to minimize the flow matching loss defined in Equation \eqref{eq:loss}. We use a learning rate of $10^{-4}$ and a batch size of $4096$. We use half of the posterior samples for training, reserving the other half for inference, or evaluating the learned harmonic mean estimator.

To evaluate the reciprocal marginal likelihood $\hat{\rho}$ via Equation \eqref{eqn:harmonic_mean_retargeted}, we require the normalized target density $\varphi(\theta_i)$. Once the velocity field is trained, we compute this density by solving the probability flow \ac{ODE}, using autodifferentiation to evaluate the divergence term. However, to ensure that the target distribution has thinner tails than the posterior, we introduce a temperature parameter $T \in (0, 1)$, which scales the variance of the flow's base distribution. This has the effect of concentrating the flow distribution, ensuring that it is contained within the posterior, as required for the learned harmonic mean estimator to have finite variance.

\section{Numerical experiments}
We perform a series of numerical experiments to validate the learned harmonic mean estimator with flow matching on a series of challenging examples with highly multimodal posteriors. We first revisit the Rastrigin example from previous work \citep{mcewen2023machine}, demonstrating the improved performance of flow matching over discrete normalizing flows. We then consider a mixture of five Gaussians in $20$ dimensions, showcasing the method's ability to handle complex posteriors in moderate dimensions.

\subsection{Rastrigin example}
In the original learned harmonic mean work \citep{mcewen2023machine}, the method's accuracy was validated on several examples, including the Rastrigin function, which is a common challenging benchmark problem for testing marginal likelihood estimators. In this work, we revisit this example with flow matching to demonstrate the improvements of the architecture in handling complex highly multimodal posteriors over discrete flows. The function is given by
\begin{equation}
	f({x}) = 10 d + \sum_{i=1}^{d} \bigl [ x_i^2 - 10 \cos ( 2 \pi x_i ) \bigr ],
\end{equation}
where $d$ denotes dimension. We use a log likelihood $\log \mathcal{L}(x) = -f({x})$ and a uniform prior with $x_i \in [-6, 6]$ for $i = 1, \dots, d$. We repeat the experiment from the original paper \citep{mcewen2023machine}, using an \ac{MCMC} implementation from the
\texttt{emcee} package \citep{emcee} to draw posterior samples in two dimensions. We draw $5,000$ samples for $80$ chains, with burn in of $2,000$ samples, yielding $3,000$ posterior samples per chain. We repeat this experiment $100$ times with different seeds. We use an architecture with $6$ layers and $256$ hidden units for the

Figure \ref{fig:rastrigin_corner} shows an example corner plot of the posterior samples used for inference in red, along with samples drawn from the flow at temperature $T=0.98$ in blue. The target distribution given by the concentrated flow is contained within the posterior as is required for the learned harmonic mean estimator. We keep the temperature high to ensure that all modes are well represented by the target, as for the Rastrigin function there are some modes present with very low weighting. All represented modes are captured by the flow accurately.

The distribution of marginal likelihood values computed by the learned harmonic mean estimator for this example are shown in Figure \ref{fig:rastrigin_violin}. We also show the mean error estimate (estimated) and compare it to the standard deviation across runs (measured). We plot the ground truth value computed by numerical integration as the red dashed line. It can be seen that the estimates obtained with the learned harmonic mean are accurate and their error estimates reliable.

\subsection{Gaussian mixture model}
\begin{figure}[ht]
	\centering
	\includegraphics[width=12cm]{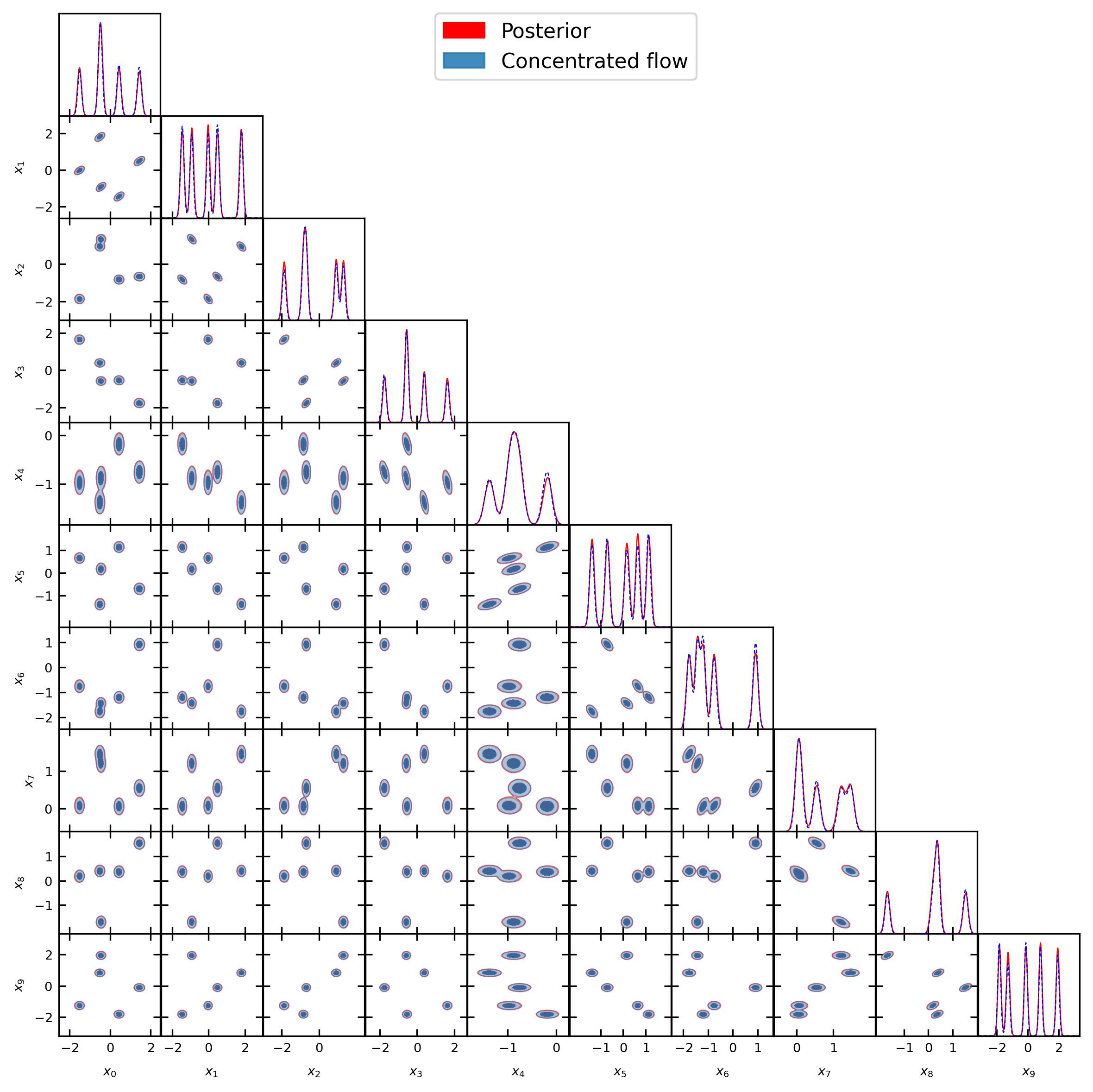}
	\caption{Corner plot for the first five dimensions of samples from the posterior (red) and trained flow with temperature $T=0.95$ (blue). The target distribution given by the concentrated flow is concentrated within the posterior, as required for the learned harmonic mean estimator. The challenging topology is correctly captured by the flow.}
	\label{fig:gm_corner}
\end{figure}

\begin{figure}[ht]
	\centering
	\includegraphics[width=10cm]{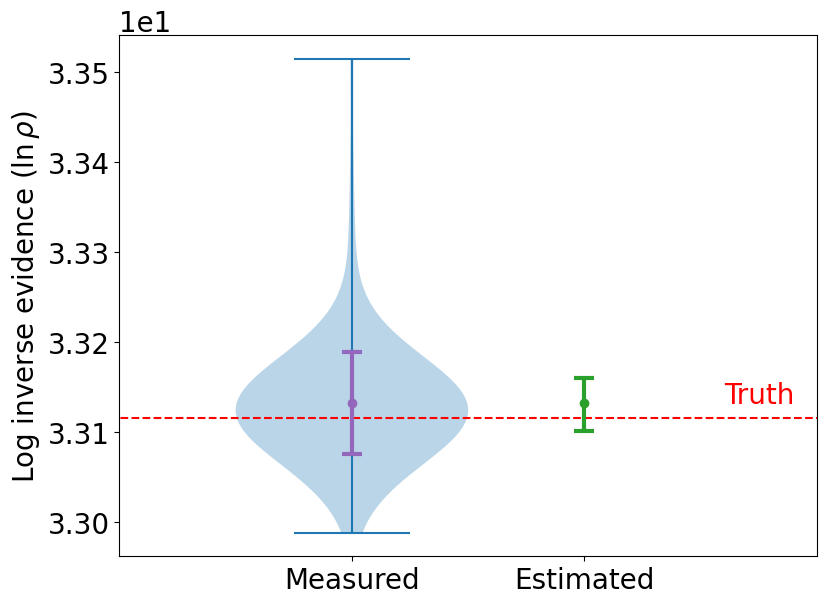}
	\caption{Marginal likelihood computed by the learned harmonic mean estimator for the mixture of five Gaussians in $20$ dimensions.
		Violin plot of $100$ runs of the experiment, showing the distribution of marginal likelihood values (measured) along with the error estimate computed by \texttt{harmonic} (estimated). The analytic ground truth is indicated by the red dashed line.}
	\label{fig:gm_violin}
\end{figure}

We also consider an example with a mixture of five Gaussian components in $20$ dimensions. The likelihood is given by
\begin{equation}
	\mathcal{L}(\mathbf{x}) = \sum_{k=1}^{K} w_k \exp \left[ -\frac{1}{2} (\mathbf{x} - \boldsymbol{\mu}_k)^\intercal \boldsymbol{\Sigma}_k^{-1} (\mathbf{x} - \boldsymbol{\mu}_k) \right],
\end{equation}
where $K=5$ denotes the number of components and $d=20$ the parameter space dimension. We consider equal weights of components $w_k = 1/K$ This model is particularly useful for validation as it is highly multimodal and in moderate dimensions, but the analytic ground truth for the evidence is still tractable, allowing for an accurate comparison. We assume a broad uniform prior. We initialised the mixture components with means vectors $\boldsymbol{\mu}_k$ distributed randomly within a $[-2, 2]^d$ hypercube. To introduce challenging correlations, the covariance matrices $\boldsymbol{\Sigma}_k$ are constructed with randomly introduced non-zero off-diagonal elements between adjacent dimensions. The scaling of the covariance matrices for each component is $0.01$, to ensure modes were non-overlapping. Since we focus on estimation of the marginal likelihood given posterior samples, rather than techniques to sample the posterior, we simply sample directly using the random module in \texttt{JAX} \citep{jax2018github}. Specifically, we draw samples from the individual mixture components, concatenate and then shuffle them. This ensures all components are well represented, as this example is a challenging case for most samplers. We generate $40,000$ samples, dividing them into $200$ chains of length $2,000$. We use an architecture with $10$ layers and $256$ hidden units for the velocity field.

We consider the same experimental setup as before. We repeat the experiment $100$ times with different seeds. Figure \ref{fig:gm_corner} shows an example corner plot for the first five dimensions of samples from the posterior in red, along with samples drawn from the flow at temperature $T=0.95$ in blue. The target distribution given by the concentrated flow is contained within the posterior, as required for the learned harmonic mean estimator.

Figure \ref{fig:gm_violin} shows the distribution of marginal likelihood values computed by the learned harmonic mean estimator for this example. As before, we also show the mean error estimate  and compare them to the standard deviation measured from the $100$ experiments (measured). We plot the analytic ground truth value as the red dashed line. It is clear that the learned harmonic mean estimator is accurate even in this very challenging scenario and its error estimators are also reasonable.

\section{Conclusion}
In this work we introduced flow matching-based models as a powerful architecture for the learned harmonic mean estimator of the marginal likelihood. This allowed us to overcome limitations of the density estimators considered previously that were based on discrete normalizing flows or bespoke classical density estimators, which struggled with highly multimodal posteriors. By leveraging the expressivity of \acp{CNF} trained with flow matching, we demonstrated that the learned harmonic mean estimator can accurately and robustly estimate the evidence for highly complex posteriors without the need for fine-tuning or heuristic modifications to the base distribution.

We validated our approach on challenging examples, including the Rastrigin function and a mixture of five Gaussians in $20$ dimensions, showcasing its ability to handle highly complex posteriors. The results confirmed that the flow matching architecture effectively captures the essential features of the posterior while ensuring that the target density remains concentrated within high-probability regions, as required for the learned harmonic mean.

In future work we plan to apply this methodology to real-world problems, where complex highly multimodal posteriors are common. This includes revisiting gravitational wave astrophysics applications. We also plan to explore refining the architecture further to be able to tackle very high-dimensional inference tasks, for instance in imaging or field-level cosmological inference.

\bibliographystyle{unsrt}
\bibliography{sources}

\begin{acronym}
	\acro{MCMC}[MCMC]{Markov chain Monte Carlo}
	\acro{HMC}[HMC]{Hamiltonian Monte Carlo}
	\acro{NF}[NF]{normalizing flow}
	\acro{CPU}[CPU]{central processing unit}
	\acro{GPU}[GPU]{graphical processing unit}
	\acro{TPU}[TPU]{tensor processing unit}
	\acro{ML}[ML]{machine learning}
	\acro{JIT}[JIT]{just-in-time}
	\acro{SBI}[SBI]{simulation-based inference}
	\acro{NUTS}[NUTS]{No-U-Turn sampler}
	\acro{VI}[VI]{variational inference}
	\acro{KL}[KL]{Kullback-Leibler}
	\acro{CNF}[CNF]{continuous normalizing flow}
	\acro{ODE}[ODE]{ordinary differential equation}
	\acro{GW}[GW]{gravitational wave}
	\acro{MLP}[MLP]{multi-layer perceptron}
\end{acronym}

\end{document}